\documentclass[pra,superscriptaddress,twocolumn,floatfig,showpacs,floatfix]{revtex4-1}
\usepackage{graphicx,color}
\usepackage{amsmath,amssymb,bm}
\usepackage{braket}		

\usepackage{mathtools}

\usepackage[plainpages=false,pdfpagelabels,colorlinks=true,linkcolor=red,urlcolor=blue,citecolor=blue,pdftitle={Eigenstate thermalization and quantum chaos in the Holstein polaron model},pdfauthor={},pdfdisplaydoctitle=true,pdfduplex=DuplexFlipLongEdge]{hyperref}

\definecolor{darkred}{rgb}{0.90,0,0}
\definecolor{darkgreen}{rgb}{0,0.60,.2}
\definecolor{darkblue}{rgb}{0,0,1}
\definecolor{grey}{cmyk}{0,0,0,0.25}
\definecolor{orange}{cmyk}{0,0.6,1,0}

\begin{document}
\title{
Eigenstate thermalization and quantum chaos in the Holstein polaron model
}

\author{David Jansen}
\affiliation{Department of Physics, Arnold Sommerfeld Center for Theoretical Physics (ASC), Ludwig-Maximilians-Universit\"{a}t M\"{u}nchen, D-80333 M\"{u}nchen, Germany}
\affiliation{Munich Center for Quantum Science and Technology (MCQST), Schellingstr. 4, D-80799 M\"{u}nchen, Germany}

\author{Jan Stolpp}
\affiliation{Institut for Theoretical Physics, Georg-August-Universit\"at G\"ottingen, D-37077 G\"ottingen, Germany}

\author{Lev Vidmar}

\affiliation{Department of Theoretical Physics, J. Stefan Institute, SI-1000 Ljubljana, Slovenia}

\author{Fabian Heidrich-Meisner}
\affiliation{Institut for Theoretical Physics, Georg-August-Universit\"at G\"ottingen, D-37077 G\"ottingen, Germany}

\begin{abstract}
The eigenstate thermalization hypothesis (ETH) is a successful theory that provides sufficient criteria for ergodicity in quantum many-body systems.
Most studies were carried out for Hamiltonians relevant for ultracold quantum gases and single-component systems of spins, fermions, or bosons.
The paradigmatic example for thermalization in solid-state physics are phonons serving as  a bath for electrons.
This situation is often viewed from an open-quantum system perspective.
Here, we ask whether a minimal microscopic model for electron-phonon
coupling is quantum chaotic and whether it obeys ETH, if viewed as a closed quantum system. Using exact diagonalization, we address this question in the framework of the 
Holstein polaron model. 
Even though the model describes only a single itinerant electron, whose coupling to dispersionless phonons is the only integrability-breaking term, we find that the spectral statistics and the structure of Hamiltonian eigenstates exhibit essential properties of the corresponding 
random-matrix ensemble.
Moreover, we verify the ETH ansatz  both for diagonal and offdiagonal matrix elements of typical phonon and electron observables, and show that the ratio of their variances equals the value predicted from random-matrix theory.
\end{abstract}


\maketitle


\section{Introduction}
\label{sec:intro}

Understanding whether and how an isolated quantum many-body system approaches thermal equilibrium after being driven far from equilibrium has been a tremendous theoretical challenge since the birth of quantum mechanics~\cite{vonneumann_29, Goldstein2010}.
This topic became viral after quantum-gas experiments~\cite{Bloch2008,Langen2015,Eisert2015} and other quantum simulators such as ion traps became available that are, to a good approximation, closed quantum systems.
Using these platforms, a series of experiments on nonequilibrium dynamics was conducted, focusing either on the generic case of thermalizing systems~\cite{Trotzky2012,Kaufman2016,Neill2016,tang_kao_18} or the exceptions such as integrable models~\cite{Kinoshita2006,gring_kuhnert_12,langen15a}
or many-body localization~\cite{Schreiber2015,Choi2016,Smith2016}.

An additional impetus to the theoretical community was given by the work of Rigol {\it et al.}~\cite{rigol_dunjko_08} who demonstrated that the eigenstate thermalization hypothesis (ETH), pioneered by Deutsch~\cite{deutsch_91} and Srednicki~\cite{srednicki_94}, provides a relevant framework to describe statistical properties of eigenstates of lattice Hamiltonians, applicable to ongoing experiments with ultracold atoms on optical lattices~\cite{Kaufman2016,Choi2016,Schreiber2015}.
As a crucial consequence, if the ETH is satisfied in a given quantum many-body system, it implies thermalization of local observables after a unitary time evolution~\cite{rigol_srednicki_12, dalessio_kafri_16,mori_ikeda_18, deutsch_18}.

So far, the ETH has been verified for a wide number of lattice models such as nonintegrable spin-1/2 chains~\cite{rigol_09a, rigol_santos_10, steinigeweg_herbrych_13, beugeling_moessner_14, kim_ikeda_14, luitz_16, luitz_barlev_16, garrison_grover_18, richter_gemmer_18, khaymovich_haque_18, hamazaki_ueda_19}, ladders~\cite{beugeling_moessner_14, steinigeweg_khodja_14, khodja_steinigeweg_15, beugeling_moessner_15} and square lattices~\cite{fratus_srednicki_15, mondaini_fratus_16, mondaini_rigol_17}, interacting spinless fermions~\cite{rigol09, neuenhahn_marquardt_12}, Bose-Hubbard~\cite{sorg14, beugeling_moessner_14} and Fermi-Hubbard chains~\cite{mondaini_rigol_15}, dipolar hard-core bosons~\cite{khatami_pupillo_13}, quantum dimer models~\cite{lan_powell_17} and Fibonacci anyons~\cite{chandaran_schulz_16}.
In these examples, mostly, direct two-body interactions in systems of either spins, fermions or bosons are responsible for rendering the system ergodic.
In condensed matter systems, however, the presence of phonons is ubiquitous.
In fact, the textbook example of thermalization in solid-state physics is the phonons being a bath for the electrons.
There, the distinction between bath and system is  not made by a real-space bipartition that is often considered in ETH studies,
but by tracing out one type of physical degree of freedom, namely the phonons.
Moreover, in metals, direct electron-electron interactions are often irrelevant while the coupling to phonons leads to effective interactions  responsible for thermalization and transport. 

A second timely motivation to study nonequilibrium problems in electron-phonon coupled systems stems from the experimental advances with time-resolved spectroscopy~\cite{giannetti_capone_16}. Such experiments  have stimulated an increased interest in exploring nonequilibrium dynamics of isolated electron-phonon coupled systems theoretically~\cite{ku2007,vidmar11,fehske11,vidmar11c,golez12b,li13,hohenadler13,kemper13,sentef13, murakami_werner_15,dorfner_vidmar_15,werner_eckstein_15,brockt_dorfner_15,kogoj_vidmar_16,kogoj_mierzejewski_16, huang_chen_17,hashimoto_ishihara_17,chen_borrelli_17,huang_hoshina_19}.
Among them, systems with a single excited electron coupled to phonons (the so-called polaron case) provide a platform for accurate numerical simulations~\cite{bonca99,fehske2007,vidmar11,dorfner_vidmar_15,brockt_dorfner_15}, allowing us to demonstrate,  for certain nonequilibrium protocols, equilibration~\cite{golez12b,dorfner_vidmar_15} and indications for thermalization~\cite{kogoj_vidmar_16}.

\begin{figure}[b]
\includegraphics[width=1.00\columnwidth]{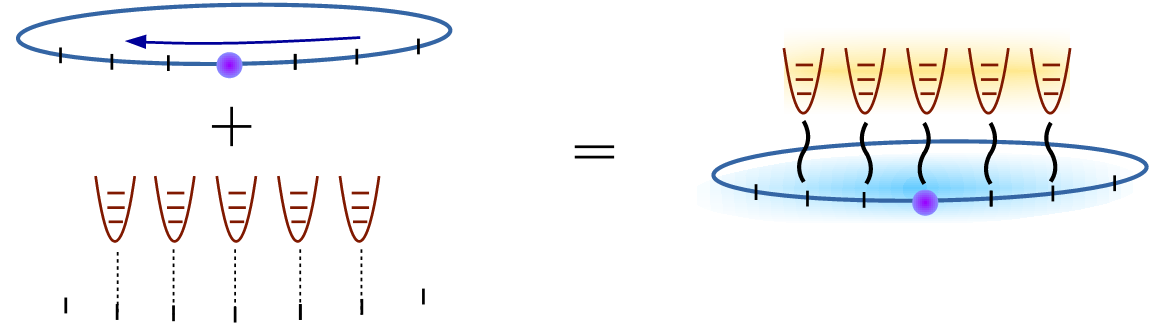}
\caption{
Quantum chaos and thermalization (right panel) emerging from coupling of a single electron (upper left) to independent local harmonic oscillators (lower left).
}
\label{fig:sketch}
\end{figure}

In this paper, we explore quantum ergodicity of a polaron system by studying indicators of quantum chaos and the validity of the ETH, which implies thermalization for generic far-from-equilibrium initial states.
We focus on the one-dimensional Holstein polaron model, which describes properties of a single electron locally coupled to dispersionless phonons, as sketched in Fig.~\ref{fig:sketch}.
Using an exact-diagonalization analysis, we demonstrate  that despite the apparent simplicity of the model, it exhibits clear manifestations of quantum chaos and the Hamiltonian eigenstates in the bulk of the spectrum obey the ETH.
Therefore, the model appears to  encompass  minimal ingredients for thermalization in condensed matter systems including both electrons and phonons.
Note that dispersionless phonons do not match the standard properties expected for a thermal reservoir~\cite{devega_alonso_17} since they exhibit, e.g., infinite autocorrelation times.
The results of the present study, however, suggest that the presence of phonons with a dispersion is not a requirement for thermalization and quantum ergodicity in many-body quantum systems.

We first verify that the statistics of neighboring energy-level spacings is well described by the Wigner-Dyson distribution, and that the averages of restricted gap ratios agree with predictions of the Gaussian orthogonal ensemble (GOE).
We then study matrix elements of observables in both the electron and phonon sector.
For the diagonal matrix elements we use the most restrictive criterion to test the ETH~\cite{kim_ikeda_14,mondaini_fratus_16}, i.e., we show that eigenstate-to-eigenstate fluctuations vanish for all eigenstates in a finite energy-density window.
We then present an extensive analysis of the offdiagonal matrix elements and extract the universal function at different energy densities.
Finally, we study variances of both diagonal and offdiagonal matrix elements in narrow microcanonical windows and find that their ratio approaches a universal number, which matches predictions of the GOE.

An interesting aspect of the Holstein polaron model is that the electron-phonon coupling is the only mechanism that breaks integrability of the uncoupled electronic and bosonic subsystems and gives rise to quantum ergodicity.
Since there is only a  single electron in the system, expectation values of the electron-phonon coupling energy are not extensive but of order ${\cal O}(1)$ in every Hamiltonian eigenstate.
The question of the minimal perturbation strength to render  an integrable system quantum chaotic is highly nontrivial and can be traced back to the pioneering work on ETH by Deutsch~\cite{deutsch_91}.
Recently, a large number of studies addressed the influence of integrability-breaking static impurities on quantum ergodicity and thermalization (see, e.g., Refs.~\cite{bertini_fagotti_16, fagotti_17, bastianello_deluca_18b, bastianello_deluca_18, bastianello_19}), and  showed  that an ${\cal O}(1)$ integrability-breaking term is enough to induce quantum-chaotic statistics of energy levels~\cite{santos_04, barisic_prelovsek_09, santos_mitra_11, torresherrera_santos_14, brenes_mascarenhas_18}.
Here, we provide numerical evidence that an ${\cal O}(1)$-integrability-breaking term, introduced by an itinerant electron coupled to noninteracting phonons, is sufficient to observe perfect ETH properties of all Hamiltonian eigenstates in the bulk of the spectrum.

The paper is organized as follows.
We introduce the Holstein polaron model and its basic spectral properties in Sec.~\ref{sec:Holpol}.
In Sec.~\ref{sec:qchaos} we study quantum-chaos indicators in the statistics of the neighboring level spacings for different parameter regimes of the model.
We then focus on the analysis of the ETH in Sec.~\ref{sec:eth} and explore both diagonal and offdiagonal matrix elements of observables.
We conclude in Sec.~\ref{sec:conclusion}.

\section{The Holstein polaron model}
\label{sec:Holpol}

The one-dimensional Holstein polaron model on a lattice with $L$ sites is described by the Hamiltonian
\begin{equation} \label{def_Hpol}
\hat H = \hat H_{\rm kin} + \hat H_{\rm ph} + \hat H_{\rm eph} \, .
\end{equation}
It consists of the electron kinetic energy operator
\begin{equation} \label{def_Hkin}
\hat H_{\rm kin} = - t_0 \sum_{j=1}^L \left( \hat c_j^{\dag} \hat c_{j+1}^{\phantom{\dag}} + \hat c_{j+1}^{\dag} \hat c_j^{\phantom{\dag}} \right) \, ,
\end{equation}
where $\hat c_j$ is the electron annihilation operator acting on site $j$ (in the Holstein model, a spin-polarized gas of electrons is considered).
Next, the phonon-energy operator reads
\begin{equation} \label{def_Hph}
\hat H_{\rm ph} = \hbar \omega_0 \sum_{j=1}^L \hat b_j^{\dag} \hat b_j^{\phantom{\dag}} \, ,
\end{equation}
where $\hat b_j$ is the phonon annihilation operator at site $j$ and $\hbar \equiv 1$,
and the local coupling of phonons to the electron density $\hat n_j = \hat c_j^{\dag} \hat c_j^{\phantom{\dag}}$ is
\begin{equation} \label{def_Heph}
\hat H_{\rm eph} = \gamma \sum_{j=1}^L \left( \hat b_j^{\dag} + \hat b_j^{\phantom{\dag}} \right) \hat n_j \, .
\end{equation}
We use periodic boundary conditions, $\hat c_{L+j} \equiv \hat c_j $ and we set the lattice spacing to $a=1$.
Throughout the paper we use $t_0$ as the energy unit and set $t_0=1$ in numerical calculations.

\begin{figure}[!]
\includegraphics[width=0.99\columnwidth]{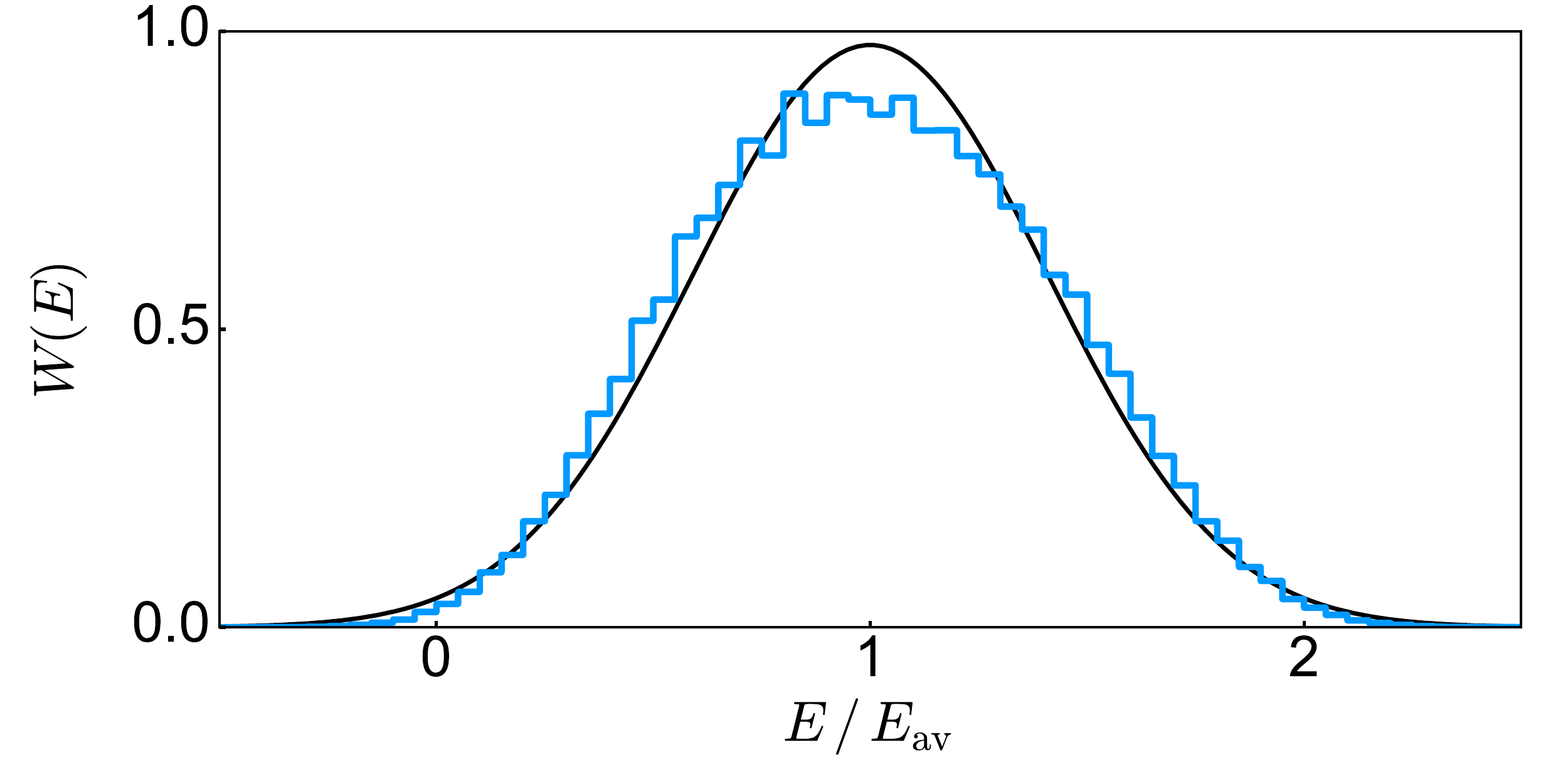}
\caption{
The density of eigenstates $W(E)$ of the Holstein polaron model using a bin width $\Delta E = E_{\rm av}/20$.
The system parameters are $\omega_0/t_0=1/2$, $\gamma/t_0=1/\sqrt{2}$, $L=8$ and $M=3$, which corresponds to ${\cal D} = L(M+1)^L \approx 5 \times 10^5$ eigenstates.
The smooth line is a Gaussian function
[see Eq.~\eqref{eq:gauss}], where $\Gamma$ is given by Eq.~(\ref{variance}).
}
\label{fig:dos}
\end{figure}

We employ full exact diagonalization to numerically obtain all  eigenvalues and eigenstates of the Holstein polaron model, denoted by $\{ E_\alpha \}$ and $\{ |\alpha\rangle \}$, respectively.
We truncate the number of local bosonic degrees of freedom, with  $M$ representing the maximal number of phonons per site.
We exploit translational invariance to split the Hamiltonian into $L$ distinct sectors with quasimomenta $k$.
Each sector contains one electron, i.e.,
$\langle \hat N \rangle = \langle \sum_j \hat c_j^\dagger \hat c_j \rangle = 1$.
The Hilbert-space dimension of each $k$ sector is ${\cal D}' = (M+1)^L$, and hence the total Hilbert-space dimension is ${\cal D} = L (M+1)^L$.

Since we use exact diagonalization, it is necessary to deal with the  finite cutoff $M$ in the local phonon number.
This introduces a dependence on $M$ in some quantities.
The approach taken in this paper is to establish, for a fixed $M$, quantum-chaotic properties of the model and the validity of the ETH for eigenstates that fall within a finite window of energy densities.
We will argue that the qualitative conclusions do not depend on the choice of $M$.

We define the distribution of the Hamiltonian eigenvalues $\{E_\alpha\}$, i.e., the density of eigenstates, as
\begin{equation}
 W(E) = {\cal D}^{-1} \sum_{\alpha=1}^{\cal D} \delta (E-E_\alpha) \, .
\end{equation}
In Fig.~\ref{fig:dos}, we plot $W(E)$ for a system with $L=8$, $M=3$ and the model parameters $\gamma/t_0=1/\sqrt{2}$, $\omega_0/t_0=1/2$.
The density of eigenstates is close to a Gaussian function 
\begin{equation}
g(E) = E_{\rm av} /(\sqrt{2\pi}\Gamma) e^{-(E - E_{\rm av})^2/(2\Gamma^2)}\,,
\label{eq:gauss}
\end{equation}
which is shown in the same figure (solid line).
The average energy $E_{\rm av}$ and the energy variance $\Gamma^2$ are given by
\begin{equation} \label{def_Eavr}
E_{\rm av} = \langle \hat H \rangle = {\cal D}^{-1} {\rm Tr}\{\hat H\} = L \, \frac{\omega_0 M}{2} \,
\end{equation}
and
\begin{equation} \label{variance}
 \Gamma^2 = \langle \hat H^2 \rangle - \langle \hat H \rangle^2 = L \, \frac{\omega_0^2 M (M+2)}{12} + 2 t_0^2 + \gamma^2 M \, .
\end{equation}
The latter result is consistent with the expectation that the relative standard deviation $\Gamma/L$ vanishes as $1/\sqrt{L}$ when $L \to \infty$.
On the other hand, if $L$ is kept fixed and $M \to \infty$, the relative standard deviation $\Gamma/M$ does not vanish but goes to a constant $\Gamma/M \to \omega_0 \sqrt{L/12}$.
We study finite-size dependencies by fixing $M$ and making $L$ larger.

\section{Quantum-chaos indicators}
\label{sec:qchaos}

We first study statistical properties of the spectrum of the Holstein polaron model.
A standard approach is to compare these properties with predictions of the random-matrix theory, in particular, to the GOE, which represents the appropriate symmetry class for the Holstein polaron model, i.e., models with time-reversal symmetry.
Historically, random-matrix theory was shown to provide a relevant framework to describe spectral properties of quantum systems whose classical counterparts are chaotic~\cite{bohigas_giannoni_84}.
As a consequence, all quantum systems with spectral statistics identical to the ones in random matrices are called quantum chaotic~\cite{santos_rigol_10a}.
Since there is no classical counterpart of the Holstein polaron model, we refer to the model as being quantum chaotic if its spectral statistics matches those of the GOE.
Note that the level-spacing statistics (to be studied below) can not distinguish between a completely ergodic system and a system with a nonzero number of nonergodic eigenstates whose fraction vanishes in the thermodynamic limit~\cite{kormos_collura_17, turner_michailidis_18a, turner_michailidis_18b, ho_choi_19, khemani_laumann_19, robinson_james_19, lin_motrunich_19, iadecola_znidaric_19}.
In Sec.~\ref{sec:eth} we show that the system under investigation here is indeed completely ergodic, i.e., all eigenstates away from the edges of the spectrum exhibit ETH properties~\cite{dalessio_kafri_16}.

In finite systems, quantum-chaotic properties are usually observed most unambiguously at intermediate parameter regimes, i.e., when all parameters are of the same order.
Moreover, the Holstein polaron model has two integrable limits (the noninteracting limit $\gamma\to 0$ and the small-polaron limit $t_0\to 0$), where the quantum-chaotic properties are expected to be absent.
Exploring the parameter regimes in which the model exhibits quantum-chaotic properties is one of the main goals of this section.

In the parameter regime where the model is quantum chaotic, this manifests itself for eigenstates at a nonzero energy density above the ground state and below the highest energy eigenstate~\cite{dalessio_kafri_16}.
In the Holstein polaron model, the ground-state energy density in the limit $L\to\infty$ is $E_{\rm min}/E_{\rm av} = 0$ and the highest-eigenstate energy density is $E_{\rm max}/E_{\rm av} = 2$.
We introduce a control parameter $\eta \in [0,1)$ to consider eigenstates in a finite energy-density window, such that
\begin{equation}
\frac{E_{\rm av} - E_\alpha}{E_{\rm av} - E_{\rm min}} < \eta \;\; \mbox{if} \;\; E_\alpha < E_{\rm av}
\end{equation}
and
\begin{equation}
\frac{E_\alpha - E_{\rm av}}{E_{\rm max} - E_{\rm av}} < \eta \;\; \mbox{if} \;\; E_\alpha > E_{\rm av} \, .
\end{equation}
In our model the corresponding set of eigenstates within the target energy-density window is
\begin{equation} \label{def_Zeta}
 {\cal Z}_\eta^{\{k\}} = \{ |\alpha\rangle \; ; \; E_\alpha / E_{\rm av} \in [1-\eta, 1+\eta] \} \, ,
\end{equation}
where $\{ k \}$ denotes the momentum sectors from which eigenstates are selected.
It is well known~\cite{santos_rigol_10b, dalessio_kafri_16} that quantum-chaos indicators can only be observed in statistics of Hamiltonian eigenvalues of a single symmetry sector.
We therefore limit our analysis in this section to the quasimomentum sector $k=2\pi/L$ in the set of eigenstates ${\cal Z}_\eta^{\{k\}}$ in Eq.~(\ref{def_Zeta}) and fix $\eta=2/3$.

\begin{figure}[!]
\includegraphics[width=0.99\columnwidth]{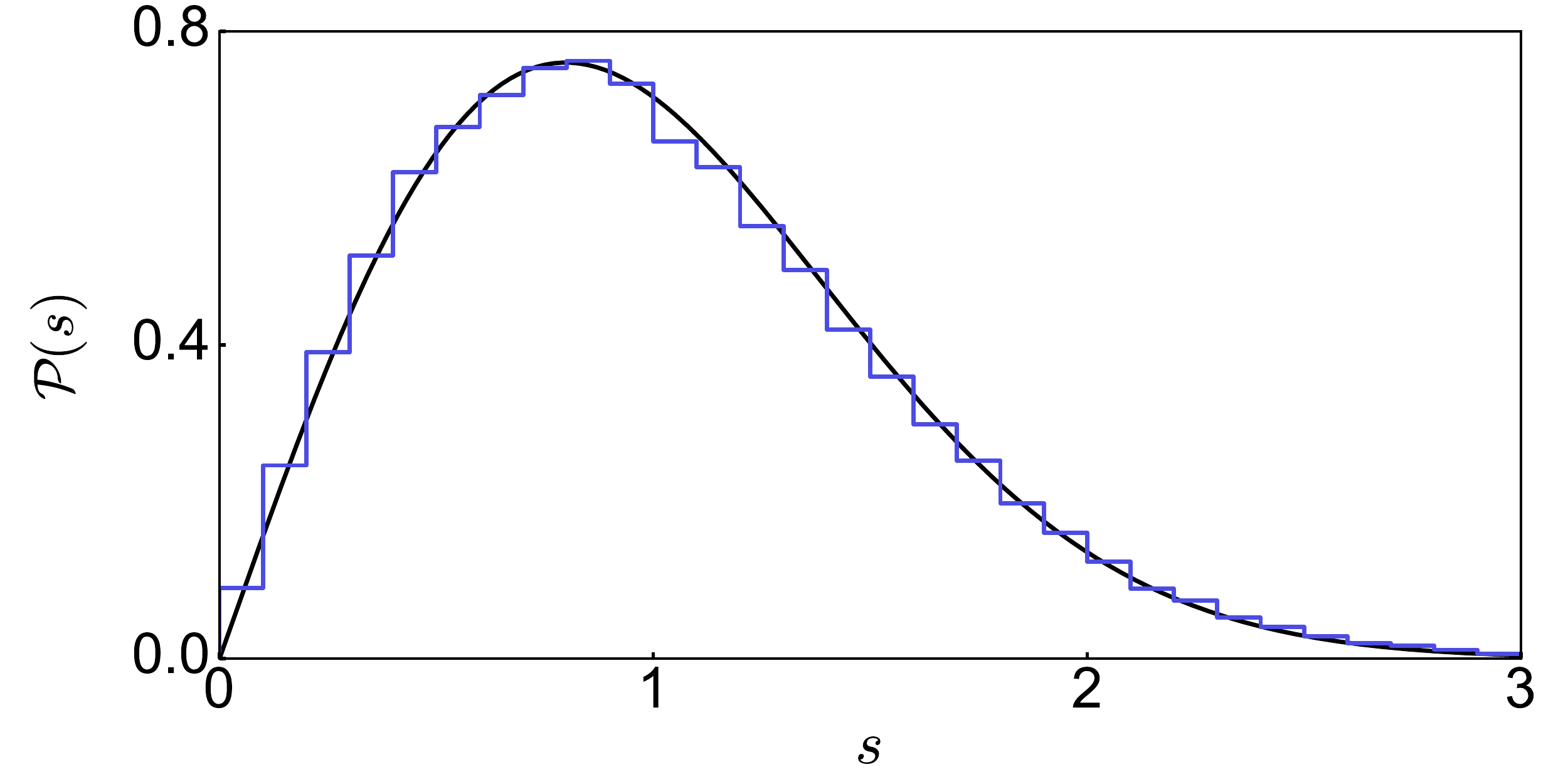}
\caption{
Level-spacing statistics ${\cal P}(s)$ after the unfolding procedure (see the text for details) for $L=8$, $M=3$, $\omega_0/t_0=0.5$ and $\gamma/t_0=1/\sqrt{2}$.
Numerical results are shown as a histogram for the $k=2\pi/L$ momentum sector.
We consider eigenstates within an energy density window given by $\eta=2/3$ in Eq.~(\ref{def_Zeta}), which for the particular model parameters corresponds to $90\%$ of all eigenstates.
Smooth lines are the Wigner-Dyson distribution ${\cal P}_{\rm WD}(s)$, given by Eq.~(\ref{def_wd}).
}
\label{fig:levelstat}
\end{figure}

We study quantum-chaos indicators by analyzing the statistics of neighboring energy-level spacings $\delta_\alpha= E_{\alpha+1}-E_{\alpha}$.
We perform an unfolding of the energy levels in ${\cal Z}_\eta^{\{k\}}$ such that the mean level spacing is one~\cite{brody_flores_81, santos_rigol_10a}.
The unfolding procedure is carried out by introducing a cumulative spectral function ${\cal G}(E) = \sum_{\alpha} \Theta(E-E_\alpha)$, where $\Theta$ is the unit step function.
In the practical analysis, we find it useful to smoothen it by fitting a polynomial of degree six $g_6(E)$ to ${\cal G}(E)$.
The spectral analysis is then performed on the unfolded neighboring level spacings $s_\alpha = g_6(E_{\alpha+1}) - g_6(E_\alpha)$.
We verified that using polynomials $g_n(E)$ of different degrees has a negligible influence on the results.

The distribution ${\cal P}(s)$ of the unfolded neighboring level spacings $s_\alpha$ for the Holstein polaron model at $\omega_0/t_0=1/2$ and $\gamma/t_0 = 1/\sqrt{2}$ is shown as a histogram in Fig.~\ref{fig:levelstat}.
We compare the distribution ${\cal P}(s)$ to the Wigner-Dyson distribution (smooth line in Fig.~\ref{fig:levelstat})
\begin{equation} \label{def_wd}
 {\cal P}_{\rm WD}(s) = \frac{\pi s}{2} e^{-\pi s^2/4} \, ,
\end{equation}
which was derived for an ensemble of $2 \times 2$ random matrices and, despite not being exact, is now considered as a prototypical distribution of the neighboring level spacings in the GOE~\cite{dalessio_kafri_16}.
We observe a reasonable agreement of ${\cal P}(s)$ with ${\cal P}_{\rm WD}(s)$.
The agreement may appear surprising since the only integrability-breaking term in the Holstein polaron model is $\hat H_{\rm eph}$~(\ref{def_Heph}), which is intensive.
Moreover, our observations  are qualitatively consistent with previous studies of level statistics in spin-1/2 chains with integrability-breaking terms of the same order~\cite{santos_04, barisic_prelovsek_09, santos_mitra_11, torresherrera_santos_14, brenes_mascarenhas_18}.

\begin{figure}[!]
\includegraphics[width=0.99\columnwidth]{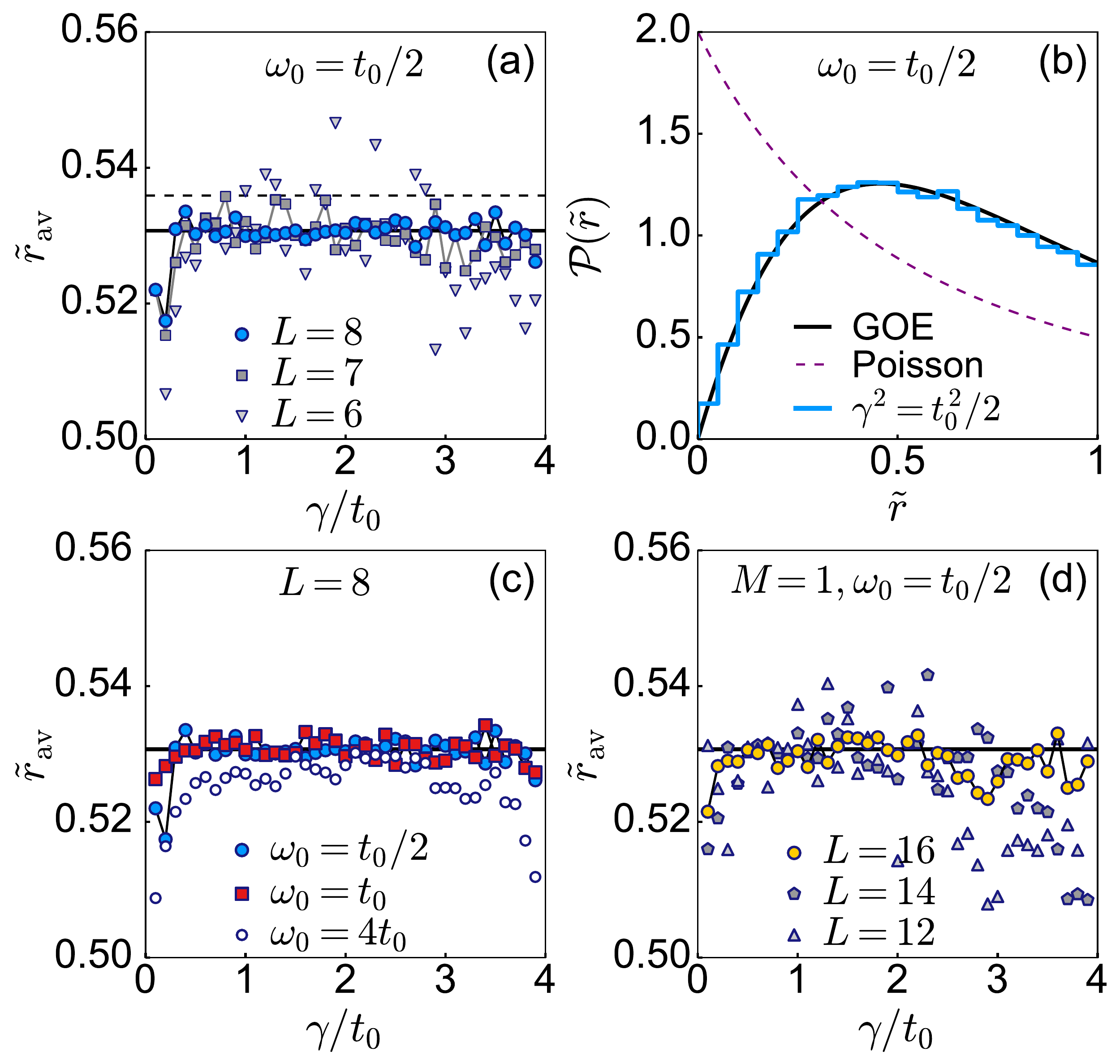}
\caption{
Statistics of the restricted gap ratio $\tilde r_\alpha$ introduced in Eq.~(\ref{def_rqrilde}).
Symbols in (a), (c) and (d) show the average $\tilde r_{\rm av}$ as a function of $\gamma/t_0$.
We show results at $\omega_0/t_0=1/2$, $M=3$ and three different system sizes $L$ in (a), results at $L=8$, $M=3$ and three different values of $\omega_0/t_0$ in (c), and results at $\omega_0/t_0=1/2$, $M=1$ and for three different system sizes $L$ in (d).
Solid lines in (a), (c) and (d) denote $\tilde r_{\rm GOE} \approx 0.5307$, which is obtained numerically for asymptotically large systems~\cite{atas_bogomolny_13}, while the dashed line in (a) denotes the analytical result $\int_0^1 \tilde r {\cal P}_{\rm GOE}(\tilde r) {\rm d}\tilde r = 4-2\sqrt{3} \approx 0.5359$ obtained from the GOE distribution ${\cal P}_{\rm GOE}(\tilde{r})$ for $3 \times 3$ matrices, see Eq.~(\ref{def_pgoe}).
(b) Histogram of the distribution ${\cal P}(\tilde{r})$ at $\omega_0/t_0=1/2$, $\gamma/t_0=1/\sqrt{2}$, $L=8$ and $M=3$.
The solid line is ${\cal P}_{\rm GOE}(\tilde{r})$ and the dashed line is the distribution $2/(1 + \tilde{r})^2$ obtained using Poisson statistics~\cite{atas_bogomolny_13}.
}
\label{fig:levelstat2}
\end{figure}

The numerical data shown  in Fig.~\ref{fig:levelstat} are for the largest numerically available system at $M=3$, and in a regime where the model parameters are of the same order.
We now extend our analysis to other parameter regimes and system sizes.
Instead of the full distribution ${\cal P}(s)$ we study a single number (defined below) derived from  the distribution.

The authors of Ref.~\cite{oganesyan_huse_07} considered the ratio of two consecutive level spacings (shortly, the gap ratio)
\begin{equation} \label{def_rq}
  r_\alpha= \frac{\delta_\alpha}{\delta_{\alpha-1}} \, ,
\end{equation}
and introduced the restricted gap ratio
\begin{equation} \label{def_rqrilde}
  \tilde{r}_\alpha = \frac{\min \{\delta_\alpha, \delta_{\alpha-1} \}} {\max \{\delta_\alpha, \delta_{\alpha-1}\}} = \min \{ r_\alpha, r_\alpha^{-1} \} \, .
\end{equation}
A convenient property of $\tilde{r}_\alpha$ is that no unfolding is necessary to eliminate the influence of a finite-size dependence through the local density of states.

We study $\tilde r_{\rm av} = \langle \tilde r \rangle_\eta$, which represents the average value of the restricted gap ratio $\tilde{r}_\alpha$ within the set of eigenstates ${\cal Z}_\eta^{k}$ introduced in Eq.~(\ref{def_Zeta}).
A numerically accurate prediction for the average of $\tilde r_\alpha$ in the GOE is $\tilde r_{\rm GOE} \approx 0.5307$~\cite{atas_bogomolny_13}.
This value is different from the average value of uncorrelated energy levels (relevant for spectra of integrable Hamiltonians with a Poisson distribution of nearest level spacings), which is $\tilde{r}_{\rm uncorr} \approx 0.3863$.
The fact that $\tilde r_{\rm GOE} > \tilde r_{\rm uncorr}$ can be understood from the level repulsion of the GOE, which suppresses the presence of  small values of $\tilde r_\alpha$ in the probability distribution ${\cal P}(\tilde r)$ 
compared to the  spectra with uncorrelated level spacings (dashed line), as illustrated in Fig.~\ref{fig:levelstat2}(b)~\cite{dalessio_kafri_16}.

Figure~\ref{fig:levelstat2}(a) shows $\tilde{r}_{\rm av}$ at $\omega_0/t_0 = 1/2$ as a function of $\gamma/t_0$ for different system sizes $L$.
When increasing $L$, the fluctuations of the averages decrease, and at $L=8$, the averages approach the GOE prediction $\tilde r_{\rm GOE}$ with high accuracy.
Deviations from $\tilde r_{\rm GOE}$ are the largest in the limits $\gamma/t_0 \to 0$ and $\gamma/t_0 \to \infty$, which is in agreement with observations in other quantum-chaotic Hamiltonians~\cite{mondaini_rigol_17}.

Before exploring the behavior of $\tilde r_{\rm av}$ in other parameter regimes, we verify that $\tilde r_{\rm av}$ does not only agree with predictions of the GOE, but that the entire distribution ${\cal P}(\tilde r)$ of $\tilde r_\alpha$ does.
We plot ${\cal P}(\tilde r)$ in Fig.~\ref{fig:levelstat2}(b) at $\omega_0/t_0=1/2$ and $\gamma/t_0 = 1/\sqrt{2}$.
The data show an excellent agreement with the analytical GOE prediction
\begin{equation} \label{def_pgoe}
{\cal P}_{\rm GOE}(\tilde r) = \frac{27}{4} \frac{\tilde r + \tilde r^2}{(1+\tilde r + \tilde r^2)^{5/2}} \, ,
\end{equation}
which is exact for $3\times 3$ matrices and very accurate for asymptotically large systems~\cite{atas_bogomolny_13}.
As a side remark, note that $\tilde r_{\rm GOE} = 0.5307$ introduced above is a numerical result for asymptotically large systems, which on the third digit differs from the result for $3 \times 3$ matrices $\int_0^1 \tilde r {\cal P}_{\rm GOE}(\tilde r) {\rm d}\tilde r = 4-2\sqrt{3} \approx 0.5359$.
Our numerical results in Fig.~\ref{fig:levelstat2}(a) are accurate enough to resolve this difference.

We next study the influence of the phonon energy $\omega_0/t_0$ on quantum-chaotic properties of the model.
Figure~\ref{fig:levelstat2}(c) compares results for $\tilde r_{\rm av}$ versus $\gamma/t_0$ at $\omega_0/t_0 = 1/2$, 1 and 4.
As a general trend, fluctuations of $\tilde r_{\rm av}$ increase with $\omega_0/t_0$.
This is, in particular, evident for $\omega_0/t_0 = 4$, where the agreement with $\tilde r_{\rm GOE}$ is weaker.
In the antiadiabatic regime of the Holstein polaron model, $\omega_0 >  4t_0$, the phonon energy becomes larger than the electron bandwidth and gaps in the spectrum increase.
As a consequence, some values of $\tilde r_\alpha$ in Eq.~(\ref{def_rqrilde}) become very small and the overall value of $\tilde r_{\rm av}$ decreases.
Since finite-size effects are expected to increase with increasing gaps, we do not study quantum-chaotic properties in the antiadiabatic regime using the existing numerical method.

Finally, we study the influence of the local phonon cutoff $M$ on the restricted gap ratio.
Figure~\ref{fig:levelstat2}(d) shows results for $M=1$ (hard-core bosons) at $L=12$, 14 and 16, which leads to identical Hilbert-space dimensions as for  the results for $M=3$ and $L=6$, 7, 8, respectively, shown in Fig.~\ref{fig:levelstat2}(a).
We observe increased fluctuations when $M$ decreases.
Still, fluctuations decrease in both cases when fixing $M$ and increasing $L$, suggesting that in the limit $L \to \infty$, the model exhibits quantum-chaotic properties even in the case of hard-core bosons.

While strictly speaking, the thermodynamic limit corresponds to sending both $L, M \to \infty$, we observe the onset of quantum chaos (Sec.~\ref{sec:qchaos}) and the validity of the ETH (to be studied in Sec.~\ref{sec:eth}) even when $L\to\infty$ and $M$ fixed.
A plausible conjecture, which remains to be verified in future work, is that our key observations remain valid when $M \to \infty$.
In the following section, we study the ETH properties of the Holstein polaron model using $M=3$, which turned out to be the best compromise between allowing for  relatively large local phonon fluctuations and still large enough lattice sizes.

\section{Eigenstate thermalization}
\label{sec:eth}

\begin{figure*}[!]
\includegraphics[width=2.0\columnwidth]{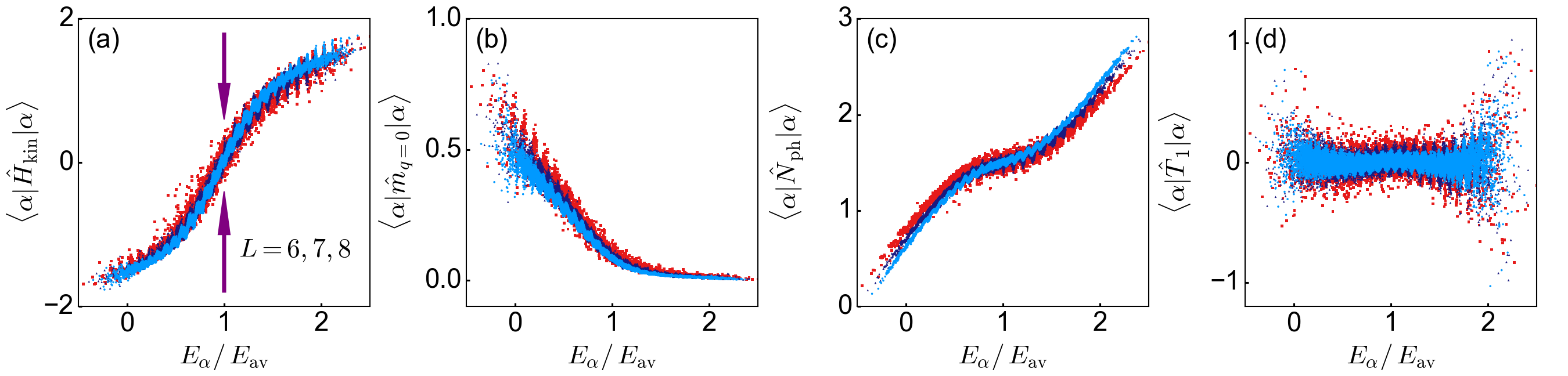}
\caption{
Diagonal matrix elements of observables $\langle \alpha | \hat O | \alpha \rangle$ for $\omega_0/t_0=1/2$, $\gamma/t_0=1/\sqrt{2}$, and $M=3$ in the quasimomentum sector $k=2\pi/L$.
Results are plotted as a function of $E_\alpha/E_{\rm av}$, where the average energy $E_{\rm av}$ is defined in Eq.~(\ref{def_Eavr}).
Symbols from the back to the front represent results for $L=6$ (red), $L=7$ (dark blue) and $L=8$ (light blue), respectively.
Arrows in (a) indicate the trend of the width of the matrix-element distribution as $L$ increases.
}
\label{fig:eev1}
\end{figure*}

\begin{figure*}[!]
\includegraphics[width=2.0\columnwidth]{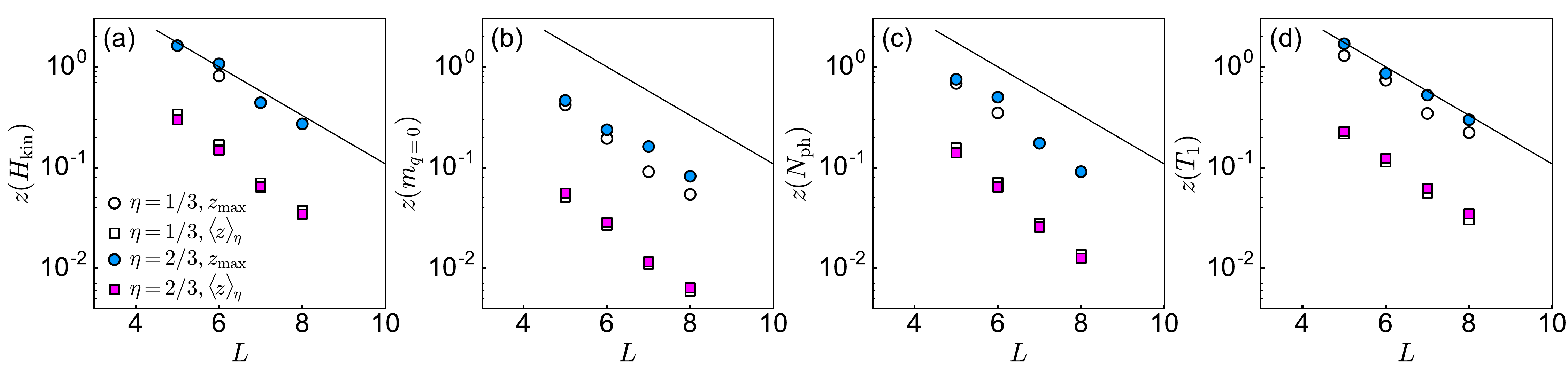}
\caption{
Statistics of eigenstate-to-eigenstate fluctuations $z_\alpha(O)$, see Eq.~(\ref{def_zfluct}), for the same observables and model parameters as in Fig.~\ref{fig:eev1}.
We show the mean $\langle z \rangle_\eta(O)$, see Eq.~(\ref{def_zmean}), and the maximum $z_{\rm max}(O)$, see Eq.~(\ref{def_zmax}).
We include eigenstates from the quasimomentum sectors with $0 \leq k \leq \pi$ that belong to the subspace ${\cal Z}_\eta$, defined in Eq.~(\ref{def_Zeta2}).
At $\eta=1/3$, the fraction of eigenstates included in ${\cal Z}_\eta$ (out of the total number of eigenstates from the quasimomentum sectors $0 \leq k \leq \pi$) are $37\%$, $44\%$, $50\%$ and $55\%$ for $L=5,6,7,8$, respectively,
while at $\eta=2/3$ the fractions are $70\%$, $79\%$, $85\%$ and $90\%$, respectively.
Solid lines are guides to the eyes, signaling an exponential decay with $L$.
}
\label{fig:eev2}
\end{figure*}

The previous section studied quantum-chaos indicators and showed that for a wide range of model parameters they agree with predictions of the GOE.
We now focus on a specific set of model parameters given by $\omega_0/t_0=1/2$, $\gamma/t_0=1/\sqrt{2}$, $M=3$, and test the validity of the ETH.

The ansatz for expectation values of observables in eigenstates of the Hamiltonian, known also as the ETH ansatz~\cite{srednicki_99}, can be written as
\begin{equation} \label{def_eth_ansatz}
 \langle \alpha | \hat O | \beta \rangle = O(\bar E) \delta_{\alpha,\beta} + e^{-S(\bar E)/2} f_O(\bar E, \omega) R_{\alpha,\beta} \, ,
\end{equation}
where $\bar E = (E_\alpha + E_\beta)/2$ is the average energy of the pair of eigenstates $|\alpha\rangle$ and $|\beta\rangle$, $\omega=E_\alpha - E_\beta$ is the corresponding energy difference, and $S(\bar E)$ is the thermodynamic entropy at energy $\bar E$.
The requirement for $O(\bar E)$ and $f_O(\bar E,\omega)$ is to be smooth functions of their arguments, while $R_{\alpha,\beta}$ are random numbers with zero mean and unit variance.
When the ETH ansatz is applicable, $O(\bar E)$ from the diagonal part is the microcanonical average of an observable $\hat O$, while the offdiagonal part can be used, e.g., to prove the fluctuation-dissipation relation for a single eigenstate~\cite{dalessio_kafri_16}.

We study four different observables, two in the electron sector and two in the phonon sector.
In the electron sector, we study the electron kinetic energy $\hat H_{\rm kin}$, defined in Eq.~(\ref{def_Hkin}), and the quasimomentum occupation operator
\begin{equation}
 \hat m_q = \frac{1}{L} \sum_{j,l=1}^L e^{-i(l-j)q} \; \hat c_j^\dagger \hat c_l \,
\end{equation}
at $q=0$.
Note that the expectation values of both electron observables are intensive quantities, which is a consequence of having only a single electron in the Holstein polaron model.
The kinetic energy $\hat H_{\rm kin}$ is the sum over all $\hat m_q$.
Therefore, if ETH behavior is observed for $\hat m_{q=0}$ it will not immediately imply ETH behavior for $\hat H_{\rm kin}$ (and vice versa).
Nevertheless, both quantities are related and it may not be surprising to see ETH behavior in both cases.

In the phonon sector, we study the average phonon number per site $\hat N_{\rm ph} = \hat H_{\rm ph}/(\omega_0 L)$, where $\hat H_{\rm ph}$ was introduced in Eq.~(\ref{def_Hph}), and the nearest-neighbor offdiagonal matrix element of the phonon one-body correlation matrix, with the underlying operator
\begin{equation}
 \hat T_1 = \frac{1}{L} \sum_{j=1}^L \left(\hat b_j^\dagger \hat b_{j+1} + {\rm H.c.} \right) \, .
\end{equation}
The observables in both sectors are chosen such that one observable is part of the Hamiltonian~(\ref{def_Hpol}) [$\hat H_{\rm kin}$ and $\hat N_{\rm ph}$] while the other one is not [$\hat m_q$ and $\hat T_1$].

\subsection{Diagonal matrix elements of observables}
\label{sec:eev1}

We first focus on the eigenstate-expectation values $\langle \alpha | \hat O | \alpha \rangle$, i.e., on the diagonal part of the ETH ansatz in Eq.~(\ref{def_eth_ansatz}).
Figures~\ref{fig:eev1}(a)-\ref{fig:eev1}(d) show results for observables in the $k=2\pi/L$ quasimomentum sector for three different system sizes $L=6,7,8$, plotted versus the eigenstate energy density $E_\alpha/E_{\rm av}$.
In all cases we observe clear manifestations of ETH, namely, eigenstate-expectation values are only functions of the energy density.
In particular, results for different $L$ suggest that the fluctuations decrease with $L$ and as a consequence, eigenstate-expectation values become smooth and sharp functions of the energy density in the limit $L \to \infty$.

The key step in validating the ETH for the diagonal matrix elements of observables is to quantify finite-size fluctuations of the results in Fig.~\ref{fig:eev1}.
We study whether (and how) eigenstate-to-eigenstate fluctuations in the bulk of the spectrum vanish when $L \to \infty$.
To this end, we use the most restrictive criterion~\cite{kim_ikeda_14} defined below.

The ETH is expected to be satisfied for the same set of eigenstates in the bulk of the spectrum for which quantum-chaos indicators were found to exhibit GOE behavior in Sec.~\ref{sec:qchaos}.
In Eq.~(\ref{def_Zeta}), we defined these sets of eigenstates ${\cal Z}_\eta^{\{k\}}$, where $\eta$ defines the interval of energy densities and $\{ k \}$ denotes the set of quasimomentum sectors, from which eigenstates are selected.
For the analysis of eigenstate-expectation values, we include eigenstates from quasimomentum sectors $0 \leq k \leq \pi$ and denote this set of eigenstates as
\begin{equation} \label{def_Zeta2}
 {\cal Z}_\eta =  \bigcup\limits_{0 \leq k \leq \pi} {\cal Z}_\eta^{\{k\}} \, .
\end{equation}
Eigenstates from quasimomentum sectors $k < 0$ are excluded from ${\cal Z}_\eta$ since the eigenstate-expectation values at $k$ and $-k$ are identical due to time-reversal symmetry.
We introduce a measure of eigenstate-to-eigenstate-fluctuations of diagonal expectation values
\begin{equation} \label{def_zfluct}
 z_\alpha(O) = \langle \alpha+1 | \hat O | \alpha+1 \rangle - \langle \alpha | \hat O | \alpha \rangle
\end{equation}
and study two statistical properties of $z_\alpha(O)$, the mean and the maximum.
The mean is defined as
\begin{equation} \label{def_zmean}
 \langle z\rangle_\eta (O) = ||{\cal Z}_\eta||^{-1} \sum_{|\alpha\rangle \in {\cal Z}_\eta} | z_\alpha(O) | \, ,
\end{equation}
where $||{\cal Z}_\eta||$ is the number of eigenstates in ${\cal Z}_\eta$,
and the maximum is
\begin{equation} \label{def_zmax}
 z_{\rm max}(O) = \max_{|\alpha\rangle \in {\cal Z}_\eta} | z_\alpha(O) | \, .
\end{equation}

The statistics of eigenstate-to-eigenstate fluctuations, Eq.~(\ref{def_zfluct}), was first studied in Ref.~\cite{kim_ikeda_14}.
The number of eigenstates analyzed in that work was a finite fraction of the total number of eigenstates, implying that those eigenstates correspond to infinite temperature in the thermodynamic limit.
Here, in contrast, we study the statistics of fluctuations in a window of finite energy densities (similar to Ref.~\cite{mondaini_fratus_16}), defined by the parameter $\eta$ in Eq.~(\ref{def_Zeta}).
By fixing $\eta$, the fraction $f$ of eigenstates involved in the statistics relative to the total number of eigenstates increases with $L$ and eventually approaches 1 when $L\to\infty$.
For example, at $\eta=2/3$ and $L=8$ the fraction $f$ is around 0.9 (see the caption of Fig.~\ref{fig:eev2} for details).

Figures~\ref{fig:eev2}(a)-\ref{fig:eev2}(d) show the finite-size scaling of $z_{\rm max}$ and $\langle z \rangle_\eta$ for the same observables as in Figs.~\ref{fig:eev1}(a)-\ref{fig:eev1}(d).
We show results for the set of eigenstates ${\cal Z}_\eta$ using $\eta=1/3$ and $2/3$.
Remarkably, in all cases, we observe an exponential decay of fluctuations with the system size $L$, while in fact the electronic density decreases from $1/5$ to $1/8$.
These results support the validity of the ETH in the Holstein polaron model, i.e., they suggest that {\it all} eigenstates in the bulk of the spectrum obey the ETH.

\subsection{Offdiagonal matrix elements of observables}
\label{sec:eev2}

\begin{figure}[!]
\includegraphics[width=0.99\columnwidth]{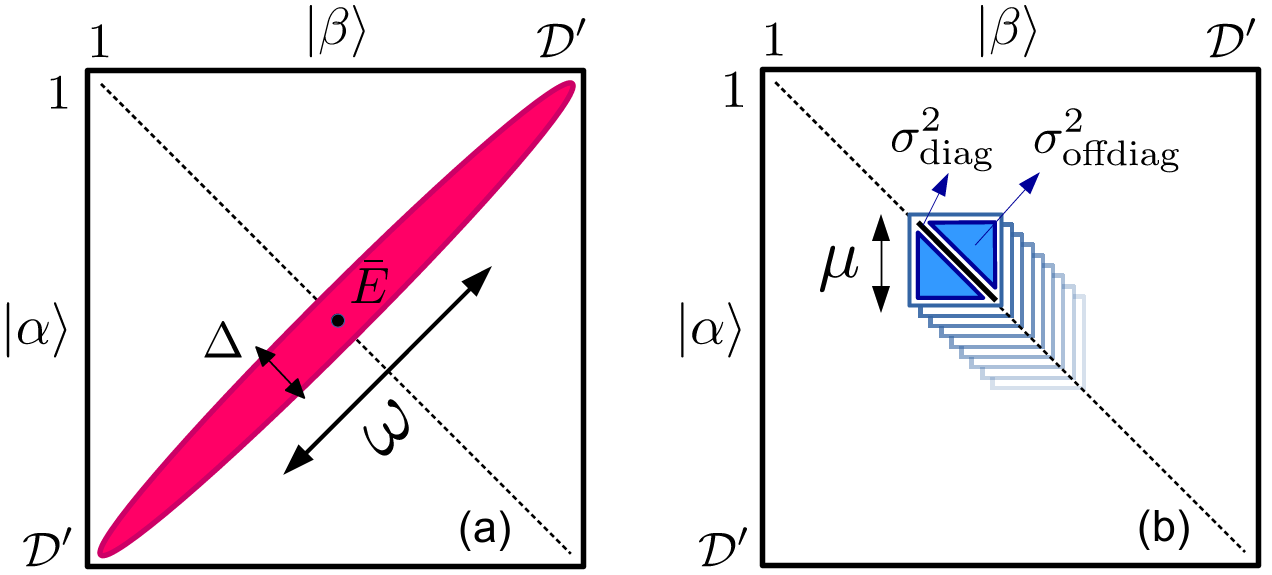}
\caption{
Sketch of a matrix of size ${\cal D}'$ with matrix elements $|\langle \alpha | \hat O | \beta \rangle|$.
The shaded region in (a) represents the matrix elements included in the sum in Eq.~(\ref{def_offdiag_avr}).
In (b), each square contains $\mu$ eigenstates, for which we calculate the variance of diagonal matrix elements $\sigma^2_{\rm diag}$, Eq.~(\ref{def_variance_diag}), and the variance of offdiagonal matrix elements $\sigma^2_{\rm offdiag}$, Eq.~(\ref{def_variance_offdiag}).
}
\label{fig:offdiag0}
\end{figure}

We now turn our focus to offdiagonal matrix elements $\langle \alpha | \hat O | \beta \rangle$ of observables and test the second part of the ETH ansatz in Eq.~(\ref{def_eth_ansatz}).
We consider the symmetry sector $k=2\pi/L$, which includes ${\cal D}'=(M+1)^L$ eigenstates.
We focus on matrix elements of pairs of eigenstates with a similar average energy $\bar E$.
An example of the set of eigenstates for which $\bar E \approx E_{\rm av}$ is sketched as a shaded region in Fig.~\ref{fig:offdiag0}(a).

\begin{figure}[!]
\includegraphics[width=0.99\columnwidth]{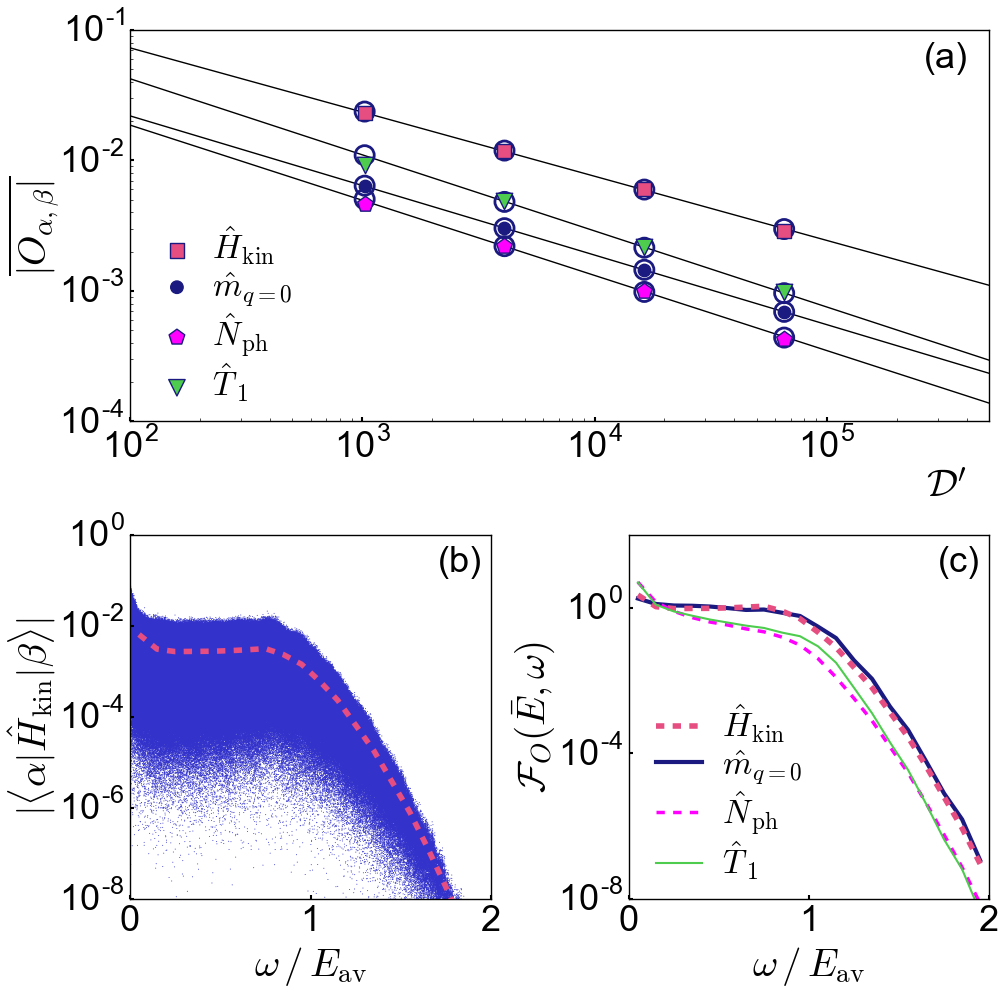}
\caption{
Offdiagonal matrix elements of observables $|\langle \alpha| \hat O | \beta \rangle|$ for $\omega_0/t_0=1/2$, $\gamma/t_0=1/\sqrt{2}$, and $M=3$ in the quasimomentum sector $k=2\pi/L$ with the Hilbert-space dimension ${\cal D}' = (M+1)^L$.
The target average energy of pairs of eigenstates is $\bar\varepsilon = \bar E/E_{\rm av} = 1$.
(a) Averages $\overline{|O_{\alpha,\beta}|}$, defined in Eq.~(\ref{def_offdiag_avr}), for system sizes $L=5,6,7,8$.
Filled symbols are obtained using $\Delta = 10^{-3}$ in Eq.~(\ref{def_offdiag_avr}) while open circles behind the filled symbols are obtained using $\Delta = 10^{-1}$.
Lines are fits to $a ({\cal D}')^{-b}$ for $L \geq 6$ and $\Delta = 10^{-3}$, where the exponent is $b=0.49$ ($\hat H_{\rm kin})$, $0.53$ ($\hat m_{q=0}$), $0.58$ ($\hat N_{\rm ph}$) and $0.58$ ($\hat T_1$).
(b) Matrix elements $|\langle \alpha| \hat H_{\rm kin} | \beta \rangle|$ versus $\omega/E_{\rm av} = (E_\alpha - E_\beta)/E_{\rm av}$.
The dashed line in (b) is a moving average in a window $\delta \omega/E_{\rm av} = 0.1$.
Lines in (c) are functions ${\cal F}_O(\bar E,\omega)$ defined in Eq.~(\ref{def_f_ma}), which are moving averages of renormalized offdiagonal matrix elements, averaged in the same window as in (b).
Results in (b) and (c)  are for $L=8$ and $\Delta=10^{-3}$.
}
\label{fig:offdiag1}
\end{figure}

Figure~\ref{fig:offdiag1}(b) shows results for the matrix elements $|\langle \alpha | \hat H_{\rm kin} | \beta \rangle|$ of the kinetic energy as a function of $\omega/E_{\rm av}$ (we only plot results for $\omega = E_\alpha - E_\beta > 0$).
We restrict the eigenstates to a narrow energy window $|(E_{\alpha} + E_{\beta}) / (2E_{\rm av}) -1|<  \Delta/2$ using $\Delta=10^{-3}$.
The results reveal strong fluctuations of $|\langle \alpha | \hat H_{\rm kin} | \beta \rangle|$, which is in agreement with the presence of the random term $R_{\alpha,\beta}$ in the ETH ansatz~(\ref{def_eth_ansatz}).
The moving average shown in the same panel indicates that the averaged function is almost flat at small $\omega / E_{\rm av}$ and then rapidly decays at larger $\omega / E_{\rm av}$, consistent with results observed in the two-dimensional transverse field Ising model~\cite{mondaini_rigol_17} and in the hard-core boson model with dipolar interactions~\cite{khatami_pupillo_13}.

\begin{figure}[!t]
\includegraphics[width=0.99\columnwidth]{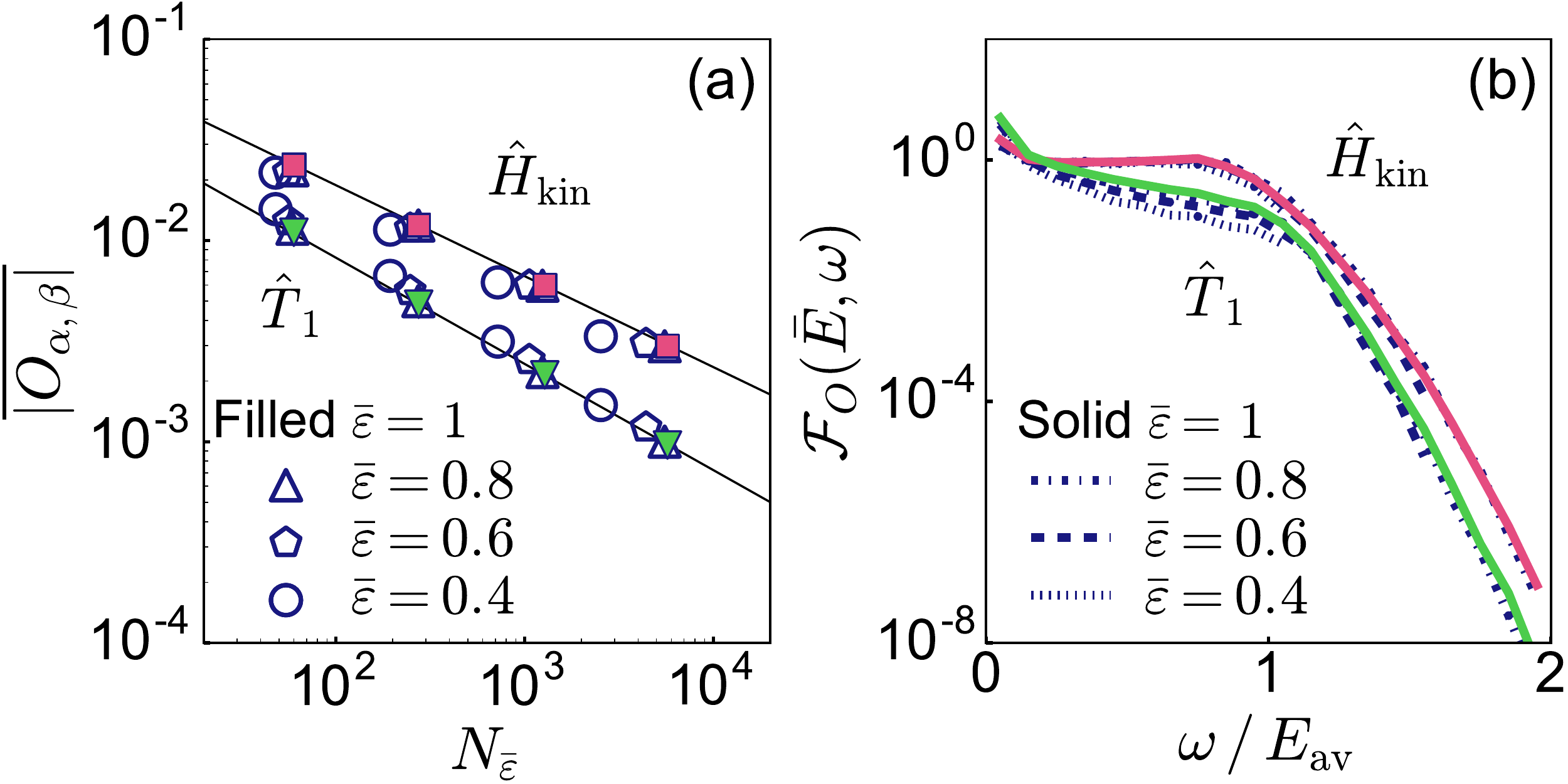}
\caption{
Offdiagonal matrix elements of observables $|\langle \alpha| \hat O | \beta \rangle|$ for the electron kinetic energy $\hat H_{\rm kin}$ and the phonon correlator $\hat T_1$, using identical system parameters as in Fig.~\ref{fig:offdiag1}.
Here, we vary target average energies of pairs of eigenstates $\bar\varepsilon = \bar E/E_{\rm av} = 1, 0.8, 0.6, 0.4$.
(a) Averages $\overline{|O_{\alpha,\beta}|}$, defined in Eq.~(\ref{def_offdiag_avr}), using $\Delta = 10^{-1}$.
Results are plotted versus the number of eigenstates $N_{\bar\varepsilon}$ within a given microcanonical window.
Lines are fits to $a (N_{\bar\varepsilon})^{-b}$ for $L \geq 6$ and $\bar\varepsilon=1$, where the exponent is $b=0.45$ ($\hat H_{\rm kin})$ and $0.53$ ($\hat T_1$).
(b) The function ${\cal F}_O(\bar E,\omega)$, defined in Eq.~(\ref{def_f_ma}), versus $\omega/E_{\rm av} = (E_\alpha - E_\beta)/E_{\rm av}$.
We use $\Delta=10^{-3}$, $L=8$ and calculate the moving average in a window $\delta \omega/E_{\rm av} = 0.1$.
}
\label{fig:offdiag2}
\end{figure}

\begin{figure*}[!]
\includegraphics[width=2.0\columnwidth]{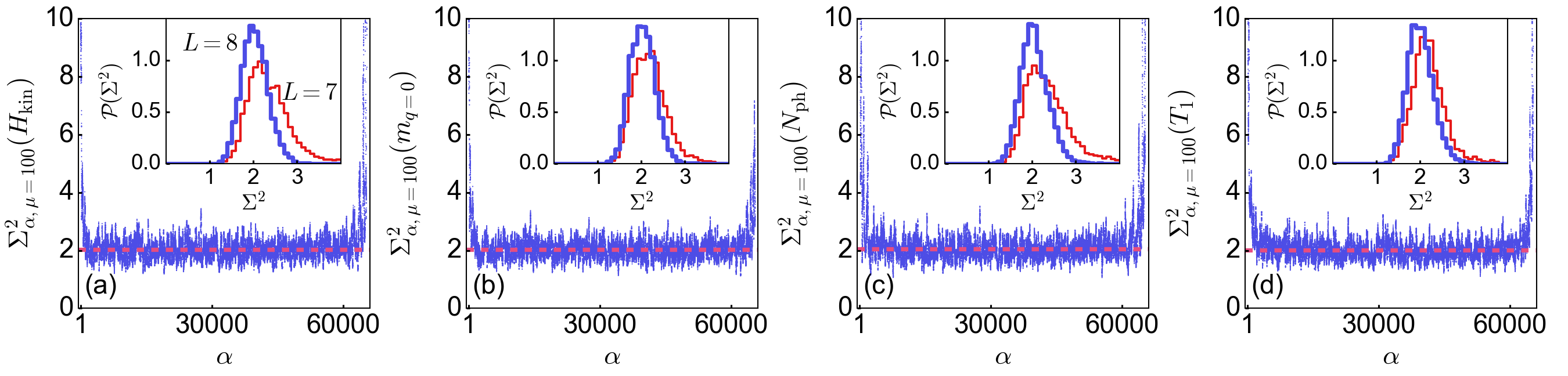}
\caption{
Variances of matrix elements of observables for $\omega_0/t_0=1/2$, $\gamma/t_0=1/\sqrt{2}$ and $M=3$ in the quasimomentum sector $k=2\pi/L$.
Main panels:
Ratio of variances $\Sigma_{\alpha,\mu}^2$, defined in Eq.~(\ref{def_ratio}), for four different observables at $L=8$.
We fix $\mu=100$ and plot $\Sigma_{\alpha,\mu=100}^2$ for every $\alpha \in [1,{\cal D}'-\mu]$.
Horizontal dashed lines are averages over those ratios of variances for which $e_{\alpha,\mu=100}/E_{\rm av} \in [1-\eta,1+\eta]$.
We use $\eta=2/3$ such that the set of eigenstates included in the average roughly corresponds to the set of eigenstates analyzed in Fig.~\ref{fig:eev2} (i.e., $90\%$ of all eigenstates).
We get $2.05, 2.04,2.07,2.03$ in panels (a)-(d), respectively.
Insets:
Histograms of $\Sigma_{\alpha,\mu=100}^2$ for lattice sizes $L=7$ and 8, using the same set of ${\alpha}$ as the one that was used to obtain the horizontal lines in the main panel.
}
\label{fig:ratio}
\end{figure*}

The overall prefactor of the matrix elements $|\langle \alpha | \hat O | \beta \rangle|$ depends on system size.
To extract that prefactor we define the average offdiagonal matrix element
\begin{equation} \label{def_offdiag_avr}
 \overline{|O_{\alpha,\beta}|} = \frac{1}{{\cal N}} \sum_{\substack{\alpha, \beta; \, \alpha \neq \beta \\ |(E_{\alpha} + E_{\beta}) / (2E_{\rm av}) - \bar \varepsilon|<  \Delta/2}} |\langle \alpha | \hat O | \beta \rangle| \, ,
\end{equation}
where the normalization ${\cal N}$ equals the number of elements included in the sum and $\bar \varepsilon = \bar E/E_{\rm av}$ is the target average energy density of the pair of eigenstates.
The ETH ansatz predicts the prefactor to be $\sim e^{-S(\bar E)/2}$, for which the leading term at $\bar\varepsilon=1$ scales as $\sim 1/\sqrt{{\cal D}'}$.
Figure~\ref{fig:offdiag1}(a) shows our numerical results for $\overline{|O_{\alpha,\beta}|}$ at $\bar\varepsilon=1$ as a function of ${\cal D}'$.
We use $\Delta=10^{-3}$ and $10^{-1}$ in Eq.~(\ref{def_offdiag_avr}) and show that results for different $\Delta$ are almost identical.
The data points can,  for all observables, accurately be fitted using the function $a({\cal D}')^{-b}$.
We find that $b$ is very close to the predicted value $b=1/2$ (see the caption of Fig.~\ref{fig:offdiag1} for the actual numerical values of $b$).
However, if one also aims at resolving eventual subleading corrections to the scaling controlled by Eq.~(\ref{def_eth_ansatz}), as discussed in Ref.~\cite{luitz_barlev_16}, more data points would be needed.

The universal properties of the offdiagonal matrix elements $|\langle \alpha | \hat O | \beta \rangle|$ encoded in $f_O(\bar E,\omega)$ can be studied after randomness and system-size dependent prefactors are removed.
We define
\begin{equation} \label{def_f_ma}
 {\cal F}_O(\bar E,\omega) = {\rm MA}\left( \frac{|\langle \alpha | \hat O | \beta \rangle| } { \overline{|O_{\alpha,\beta}|} } \right) \, ,
\end{equation}
which is a moving average of the renormalized offdiagonal matrix elements of observables.
The function ${\cal F}_O(\bar E,\omega)$ is, up to a constant prefactor, identical to the universal function $f_O(\bar E,\omega)$ used in the ETH ansatz~(\ref{def_eth_ansatz}).
We therefore plot ${\cal F}_O(\bar E,\omega)$ in Fig.~\ref{fig:offdiag1}(c) for different observables as a function of $\omega / E_{\rm av}$.
The results reveal that the offdiagonal matrix elements of both electron and phonon observables exhibit a similar scaling with $\omega$.

We complement our previous results by calculating the offdiagonal matrix elements of the pairs of eigenstates at average energy densities $\bar\varepsilon<1$.
In Fig.~\ref{fig:offdiag2}, we show results for two observables (the electron kinetic energy and the phonon correlator), while results for the other two observables studied in Fig.~\ref{fig:offdiag1}(c) exhibit a similar behavior (not shown here).
According to the ETH ansatz~(\ref{def_eth_ansatz}), the scaling of $|\langle \alpha | \hat O | \beta \rangle|$ with the system size is governed by $e^{-S(\bar E)/2}$.
In the context of Eq.~(\ref{def_offdiag_avr}), $S(\bar E)$ is the logarithm of the number of microstates $N_{\bar\varepsilon}$ in the energy density window $[\bar\varepsilon-\Delta/2,\bar\varepsilon+\Delta/2]$.
In Fig.~\ref{fig:offdiag2}(a), we plot $|\langle \alpha | \hat O | \beta \rangle|$ versus $N_{\bar\varepsilon}$ and observe a reasonably good collapse of the data for different $\bar\varepsilon$.
We fit the results to $a (N_{\bar\varepsilon})^{-b}$ and obtain $b$ close to 0.5, as predicted by the ETH ansatz.

On the other hand, the dependence of $f_O(\bar E,\omega)$ on the average energy $\bar E$ is not described by the ETH ansatz and rarely studied in the literature (see Ref.~\cite{dalessio_kafri_16} for a discussion on how $f_O(\bar E,\omega)$ is related to the fluctuation-dissipation relation).
Our results in Fig.~\ref{fig:offdiag2}(b) suggest that ${\cal F}_O(\bar E,\omega)$ in the Holstein polaron model is roughly independent of the pair energy average $\bar\varepsilon = \bar E/E_{\rm av}$.
We note that a recent study {\it ad hoc} assumed a slowly varying $f_O(\bar E,\omega)$ with $\bar E$ in a nonintegrable spin chain~\cite{richter_gemmer_18}.
Our results lend numerical support to this assumption.
The study of $f_O(\bar E,\omega)$ and its $\bar E$-dependence in other models remains an open problem.

\subsection{Variances of matrix elements of observables}

\begin{figure}[!b]
\includegraphics[width=0.99\columnwidth]{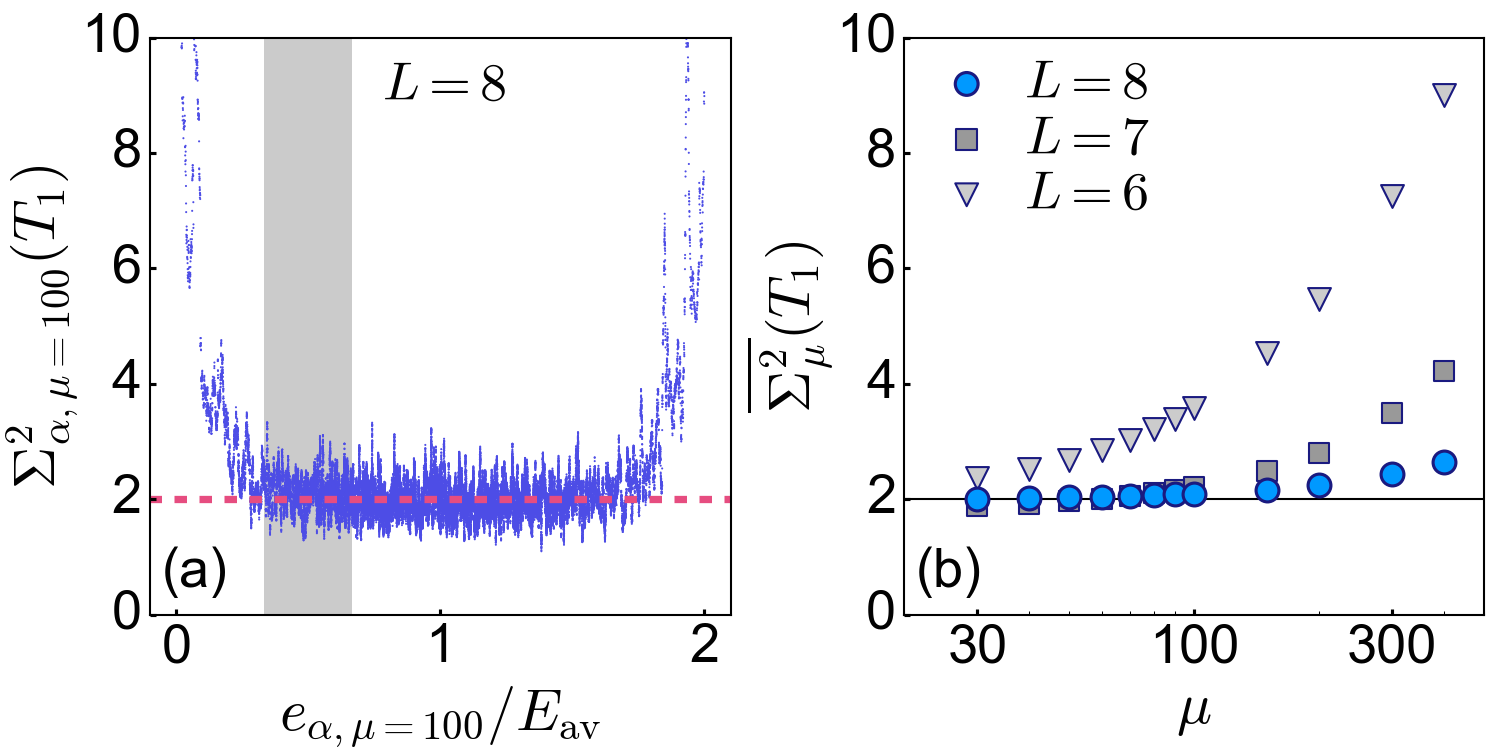}
\caption{
(a) Ratio of variances $\Sigma_{\alpha,\mu=100}^2$, defined in Eq.~(\ref{def_ratio}), for the phonon one-body correlation operator $\hat T_1$ at $L=8$ and $M=3$.
The data are identical to the ones in Fig.~\ref{fig:ratio}(d), but plotted versus the mean energy density $e_{\alpha,\mu=100}/E_{\rm av}$.
(b) Average ratio of variances $\overline{\Sigma_{\mu}^2}$, defined in Eq.~(\ref{def_avrratio}), as a function of $\mu$ for the same model parameters as in Fig.~\ref{fig:ratio}.
The average is taken over the shaded region in panel (a).
The numbers of ratios of variances included in the average are roughly $7.4 \times 10^2$, $3.0 \times 10^3$ and $1.2 \times 10^4$ for $L=6,7,8$, respectively.
}
\label{fig:ratio2}
\end{figure}

Finally, we study fluctuations of both diagonal and offdiagonal matrix elements of observables, focusing on the quasimomentum sector $k=2\pi/L$.
We define the variance of a set of consecutive diagonal matrix elements
\begin{equation} \label{def_variance_diag}
 [\sigma_{{\rm diag}}^{(\alpha,\mu)}(O)]^2 = \langle (O_{\alpha,\alpha})^2 \rangle_{\mu} - \langle O_{\alpha,\alpha} \rangle_\mu^2 \, ,
\end{equation}
where $\alpha$ determines the first matrix element and $\mu$ determines the number of matrix elements in the set.
The average $\langle O_{\alpha,\alpha} \rangle_\mu = \mu^{-1} \sum_{\rho = \alpha}^{\alpha+\mu-1} \langle \rho | \hat O | \rho\rangle$ is defined in a microcanonical window, i.e., it is an equal-weight average over $\mu$ consecutive eigenstates, starting at the index $\alpha$.
Similarly, the variance of the submatrix of offdiagonal matrix elements is defined as
\begin{equation} \label{def_variance_offdiag}
 [\sigma_{\rm offdiag}^{(\alpha,\mu)}(O)]^2 = \langle |O_{\alpha,\beta}|^2 \rangle_{\mu} - |\langle O_{\alpha,\beta} \rangle_\mu |^2 \, ,
\end{equation}
where $\langle O_{\alpha,\beta} \rangle_\mu = (\mu^2 - \mu)^{-1} \sum_{\substack{\rho, \rho' = \alpha}}^{\alpha+\mu-1} \langle \rho' | \hat O | \rho\rangle$ and $\rho \neq \rho'$.
Note that the offdiagonal matrix elements can be complex as a consequence of diagonalizing the Hamiltonian in the translationally invariant basis.
Both variances are sketched in Fig.~\ref{fig:offdiag0}(b) for a given $\alpha$ and $\mu$.
We associate the mean energy $e_{\alpha,\mu}$ to every microcanonical window defined by ($\alpha,\mu$), where $e_{\alpha,\mu} = \langle H_{\alpha,\alpha}\rangle_\mu$.

In our calculations, we fix $\mu$ and calculate the variances for every $\alpha$ in the interval $[1,{\cal D}'-\mu]$.
Our main focus is on the ratio of variances defined as
\begin{equation} \label{def_ratio}
 \Sigma_{\alpha,\mu}^2(O) = \frac{[\sigma_{\rm diag}^{(\alpha,\mu)}(O)]^2}{[\sigma_{\rm offdiag}^{(\alpha,\mu)}(O)]^2} \, .
\end{equation}
Within random-matrix theory one can show~\cite{dalessio_kafri_16} that the ratio of variances takes a universal value $\Sigma_{\rm GOE}^2(O) = 2$ for generic local observables in the GOE.
Since the ETH generalizes random-matrix theory to Hamiltonians of real physical systems, one may ask whether the same result can be obtained for matrix elements of Hamiltonian eigenstates~\cite{prosen_99} and if the answer is affirmative, at which energy densities this behavior is found.

The numerical extraction of $\Sigma_{\alpha,\mu}^2$ obeying predictions from the random-matrix theory is a nontrivial task, and the results depend on the parameters $\alpha$ and $\mu$ as well as on the Hilbert-space dimension ${\cal D}'$.
In the main panels of Figs.~\ref{fig:ratio}(a)-\ref{fig:ratio}(d), we fix $\mu=100$ and set $L=8$, $M=3$, which yields ${\cal D'} = 65536$ and hence $\mu \ll {\cal D}'$.
Our choice of $\mu$ is consistent with the one used in a recent study of a two-dimensional spin-1/2 system~\cite{mondaini_rigol_17}.
We compute $\Sigma_{\alpha,\mu=100}^2(O)$ for four different observables as a function of the eigenstate index $\alpha$ and observe $\Sigma_{\alpha,\mu=100}^2(O) \approx 2$ for the majority of eigenstates for all observables.

We quantify the excellent agreement with predictions of random-matrix theory, observed in Figs.~\ref{fig:ratio}(a)-\ref{fig:ratio}(d), by computing the averages over those ratios of variances for which $e_{\alpha,\mu=100}/E_{\rm av} \in [1-\eta,1+\eta]$ and $\eta=2/3$.
The resulting averages are shown as horizontal dashed lines in Fig.~\ref{fig:ratio} and yield, for all four observables, almost perfect agreement with $\Sigma_{\rm GOE}^2(O)$.
The insets of Figs.~\ref{fig:ratio}(a)-\ref{fig:ratio}(d) show the distributions of $\Sigma_{\alpha,\mu=100}^2(O)$  for the same set of $\alpha$ as the one that was used to obtain the horizontal lines in the main panel, and two system sizes $L=7$ and 8.
They reveal that, by increasing the system size, the distributions become narrower and their means approach the random-matrix theory result.

It is remarkable that the agreement of the ratio of variances for Hamiltonian eigenstates with the GOE predictions carries over to eigenstates away from the middle of the spectrum.
Namely, in contrast to eigenstates in the middle of the spectrum, which can be well approximated by pure states that are random superpositions of base kets in some simple basis~\cite{vidmar_rigol_17}, eigenstates away from the middle of the spectrum are no longer entirely random.
Figure~\ref{fig:ratio2}(a) shows the same data for $\Sigma_{\alpha,\mu=100}^2$ as in Fig.~\ref{fig:ratio}(d), but plotted versus the mean energy density $e_{\alpha,\mu=100}/E_{\rm av}$.
It reveals a broad region away from $e_{\alpha,\mu=100}/E_{\rm av} = 1$ where $\Sigma_{\alpha,\mu=100}^2 \approx 2$.

Finally, we study the dependence of $\Sigma_{\alpha,\mu}^2$ on $\mu$ and the system size, and we only consider eigenstates $\alpha$ away from the middle of the spectrum.
To this end, we consider a subset of ratios of variances
${\cal S}_{\mu} = \{ \Sigma_{\alpha,\mu}^2 \; ; \; e_{\alpha,\mu}/E_{\rm av} \in [1/3, 2/3] \}$,
which corresponds to a shaded region in Fig.~\ref{fig:ratio2}(a).
The average $\Sigma_{\alpha,\mu}^2$ in this region is defined as
\begin{equation} \label{def_avrratio}
 \overline{\Sigma_{\mu}^2}(O) = || {\cal S}_\mu ||^{-1} \sum_{\Sigma_{\alpha,\mu}^2 \in {\cal S}_\mu} \Sigma_{\alpha,\mu}^2(O)
\end{equation}
and plotted in Fig.~\ref{fig:ratio2}(b).
The results reveal that when $L$ increases, a larger number of eigenstates $\mu$ can be included in the variances to observe $\Sigma_{\alpha,\mu}^2 \to \Sigma_{\rm GOE}^2$.
Remarkably, this suggests that $\Sigma_{\alpha,\mu}^2 = \Sigma_{\rm GOE}^2$ in the limit $L\to\infty$, for eigenstates both in the middle and away from the center of the spectrum.

\section{Conclusions} \label{sec:conclusion}

In this paper, we addressed the question whether the hallmark features of the ETH can be observed in a paradigmatic condensed matter model that includes both electron and phonon degrees of freedom.
We studied the Holstein polaron model, which is a minimal model describing a single electron that locally couples to dispersionless phonons.
In spite of the apparent simplicity of the model, we showed that it represents an example where ETH is very well fulfilled, to a quantitative degree that is observed  in single-component models as well \cite{mondaini_rigol_17}.
In particular, we showed that:
(i) the statistics of neighboring energy-level spacings agree with predictions of the GOE;
(ii) the eigenstate-to-eigenstate fluctuations of the diagonal matrix elements of observables decay exponentially for {\it all} eigenstates in the bulk of the spectrum;
(iii) the 'universal' part of the offdiagonal matrix elements of all four observables under investigation exhibits a similar decay as a function of the energy difference of pairs of eigenstates, and exhibits an almost negligible dependence on the average energy of the pair of eigenstates; and
(iv) the ratio of variances of fluctuations of diagonal to offdiagonal matrix elements of observables fulfills predictions of the random-matrix theory.
The latter is a robust feature of Hamiltonian eigenstates also away from the middle of the spectrum.

Our results establish two important aspects of quantum many-body systems:
The first is to demonstrate that quantum ergodicity and thermalization in a many-body system of electrons and phonons does not require phonons to exhibit properties of a bath in isolation, as usually assumed in studies of open quantum systems~\cite{devega_alonso_17}.
The second is to demonstrate that an integrability breaking term of the order ${\cal O}(1)$ is sufficient to induce perfect ETH behavior.
[In our case, the eigenstate expectation value of the electron-phonon coupling operator is ${\cal O}(1)$ in every eigenstate.]
This extends previous studies~\cite{santos_04, barisic_prelovsek_09, santos_mitra_11, torresherrera_santos_14, brenes_mascarenhas_18} that showed that an ${\cal O}(1)$ integrability breaking term renders the spectral level statistics to be quantum chaotic.

There are several interesting extensions to our paper.
First, one may study generic conditions for an integrability breaking term of the order ${\cal O}(1)$ to induce ETH behavior.
Of particular interest may be systems where the uncoupled electron-phonon system is not as highly degenerate as in the Holstein model.
This can be achieved, e.g., by considering dispersive or disordered phonons.
Second, extending the analysis to finite electronic densities would be interesting, but also numerically much more challenging.
Third, exact diagonalization studies are obviously limited to small systems.
Therefore, complementary analytical insights into the problem of ETH in electron-phonon models would be desirable.
Finally, concrete studies of quench problems in the polaron case with an extensive quench energy are lacking.
This could be carried out by using, e.g., matrix-product-state (MPS) methods~\cite{schollwock2005density, schollwock2011density, paeckel_koehler_19}
(see Refs.~\cite{znidaric_scardicchio_16, kloss_barlev_18, doggen_schindler_18, hallam_morley_19} for examples of recent developments).
References~\cite{brockt_dorfner_15, schroeder_chin_16, wall_safavinaini_16, kloss_reichman_19} discuss MPS methods that are specifically designed for electron-phonon coupled systems.
These questions are left for future work.

\section*{Acknowledgments} \label{sec:ack}

We acknowledge useful discussions with J. Bon\v ca, I. de Vega, M. Goldstein, D. J. Luitz, S. Kehrein, M. Mierzejewski, P. Prelov\v sek, M. Rigol and R. Steinigeweg.
This work was supported by the Deutsche Forschungsgemeinschaft (DFG, German
Research Foundation)  under  Project No.~207383564 via Research Unit FOR 1807, and via CRC 1073 under Project No.~217133147.
D.J. acknowledges support by the Deutsche Forschungsgemeinschaft (DFG, German Research Foundation) under Germany's Excellence Strategy No.~EXC-2111-390814868.
L.V. acknowledges support from the Slovenian Research Agency (ARRS), Research Core Funding No.~P1-0044.
Part of this research was conducted at KITP at the University of California at Santa Barbara.
This research was supported in part by the National Science Foundation under Grant No.~NSF PHY-1748958.




\bibliographystyle{biblev1}
\bibliography{references}

\end{document}